# Revealing the Breakdown Mechanism and Heat Dissipation in Few-Layered semimetallic PtSe$_2$

Bubunu Biswal, Abinash Tripathy, Renu Yadav, Abhishek Misra*

Center for 2D material Research and Innovations and Department of Physics, IIT Madras, Chennai, India, 600036

*Corresponding author: Abhishek Misra:  email id: abhishek.misra@iitm.ac.in

**Abstract-** Platinum diselenide (PtSe$_2$) is an emerging two dimensional (2D) transition metal dichalcogenide with outstanding electrical, optical properties and excellent air stability. In the context of PtSe$_2$ device, understanding the high field breakdown and heat dissipation is highly essential to design the high performing as well as energy efficient devices operating at extreme limits. This work highlights the low temperature and room temperature breakdown mechanisms of PtSe$_2$ semimetal. The heat dissipation have been quantized via the interfacial thermal conductivity (ITC) of PtSe$_2$/SiO$_2$ and PtSe$_2$/$h$-BN interfaces, which has been estimated to be 14.2 MW.m$^{-2}$.K$^{-1}$ and 29.6 MW.m$^{-2}$.K$^{-1}$ respectively using Raman thermometry. We find that the device breakdown at room temperature is primarily driven by self-heating effects. In contrast, at low temperatures, the breakdown is mainly attributed to carrier multiplication in PtSe$_2$ under high electric fields, as further confirmed by Hall measurements.

**Introduction**

The heat dissipation in the next generation scaled heterogeneously stacked devices is a major bottleneck for the optimal device performances and it is severe in 2D materials due to the enhanced phonon boundary scattering and phonon confinement effect [1,2]. In the complex device structures, due to the increase in the number of interfaces in the heat disposal path causes additional thermal resistance. In the Field effect transistors (FETs), heat dissipation can occur through the source/drain contacts and the substrate. However, in the case of 2D material based devices, the higher interfacial area between the channel and the substrate dominates the contribution of the heat dissipation into the substrate and it is prominent in the case of the substrate with high thermal conductivity [3]. As the heat carrier phonons are bosonic in nature and cannot be controlled with the gate voltage, so it is crucial to design highly thermal conducting and insulating meta-surfaces based on 2D heterostructure[4]. As the thermal conductivity of an interface is different from that of the constituent materials, the understanding of interfacial thermal conductance (ITC) and thermal transport in the cross-plane direction enable us to guide the heat flow in any desired path.

The maximum operating temperature of the 2D materials stacked devices highly depends on the ITC and the adjacent materials in its surroundings. In practical device structure the ultra-thin 2D channel material is generally supported by a substrate and encapsulated with a superstrate. The hexagonal Boron Nitride ($h$-BN) has been widely used as a high-quality substrate with other 2D materials owing to its atomically smooth, dangling bond free surface [5,6]. There has been substantial enhancement by an order of magnitude of mobility of most commonly used 2D channel materials like graphene and MoS$_2$ on $h$-BN substrate compared to SiO$_2$ owing to the reduction of interfacial columbic charge impurities scattering [7]. There are numerous theoretical predictions and experimental observations demonstrating the multi-fold enhancement of ITC of 2D material interfaces with different dielectric substrates or superstrates[8,9]. Hence, choice of appropriate substrate and understanding phonon coupling and scattering at the channel–substrate interface are crucial in designing the future scaled device architectures for efficient thermal management.

The 2D topological semimetallic PtSe$_2$ has received significant research interest due to its unique thickness dependent semiconductor to Dirac semimetal transition, strong interlayer coupling, and high room temperature carrier mobility, excellent air stability and resistance to oxidation. The ballistic cross-plane phonon transport and 50-fold thermopower enhancement by band engineering in PtSe$_2$ have made it an excellent material for 2D thermoelectric devices[10]. The monolayer and bilayer PtSe$_2$ have been demonstrated as semiconducting channel material demonstrating high $I_{ON}/I_{OFF}$ of $10^9$ and mobility of 400 cm$^2$V$^{-1}$s$^{-1}$ at 10 K[11] and few layer PtSe$_2$ has been used as metallic interconnects and electrodes in next-generation sensors, photodetectors, transistors, and ultrafast photonic devices[12–15]. Despite of various application, the heat dissipation during the device operation is a major bottleneck in the ultra-scaled 2D devices. Generally, the 2D devices are fabricated on a supporting substrate like SiO$_2$, which plays a crucial role in dissipating the heat from the active device region. There is a huge scope to improve the device performances by utilizing a better heat sinking substrate through dielectric engineering. Michal et.al. have proposed to use thinner SiO$_2$ or *h*-BN as substrate to WTe$_2$ for better heat dissipation to achieve a higher threshold field of operation[16]. It has been reported that, the dangling bonds free and surface trap charge free *h*-BN substrate improves the high-temperature and high-electric-field performance of Graphene devices owing to reduction of surface optical phonon-hot carrier electron scattering[6]. Liu et.al. have derived the ITC of the PtSe$_2$ on 300nm SiO$_2$ substrate of about 8.6 MW.m$^{-2}$.K$^{-1}$,[17] whereas understanding the ITC of PtSe$_2$ interface with *h*-BN or any other suitable substrates has not been explored yet. Despite the significant potential of PtSe$_2$, its substrate dependent detailed high field breakdown and energy dissipation studies are still limited.

In this work, we explored the breakdown mechanism of PtSe$_2$ devices at low temperature and room temperature. The Current-Voltage (I-V) characteristics were thoroughly investigated at the room temperature and low temperature (5K). The room temperature breakdown is mainly due to the Joule heating of the devices, which have been understood via quantifying the heat dissipation into the substrate. The higher ITC value of PtSe$_2$/*h*-BN interface in comparison to the PtSe$_2$/SiO$_2$ interface is mainly due to the ultra-clean 2D-2D interface and the high thermal conductivity of the *h*-BN. The low temperature breakdown occurs following a super-linear current flowing in the device. This super-linear current in the device is due to the hot carrier multiplication in PtSe$_2$ which is confirmed via the Hall measurement.

**Device fabrication and measurements**

To fabricate the PtSe$_2$ two probe and four probe devices, the PtSe$_2$ bulk crystals (purchased from the supplier HQ Graphene) were exfoliated onto a thermally grown 90 nm SiO$_2$/Si substrate. The thinner and rectangular shaped few layer PtSe$_2$ flakes were targeted for the device fabrication. To fabricate the PtSe$_2$ devices on *h*-BN substrate, the *h*-BN bulk crystals were exfoliated onto the same thermally grown 90 nm SiO$_2$/Si substrate. The PtSe$_2$ flakes were picked from the substrate by the membrane assisted dry transfer technique and carefully peeled on to the *h*-BN. The PtSe$_2$/*h*-BN stacks were annealed at 200°C for 20 minutes to ensure the interaction among the interfaces. The contact electrodes are patterned with the help of electron beam Lithography and followed by the e-beam evaporation of 20 nm Titanium and 20 nm Gold deposition and lift-off process. After the electrode fabrication, all the devices were annealed for better electrode/channel contact formation. All the devices were fabricated on a single Si wafer and processed in the same run to avoid the process induced variability. The thickness of the grown SiO$_2$ is confirmed with the spectroscopic ellipsometer and the thickness of the PtSe$_2$ and *h*-BN flakes were confirmed from the Atomic Force Microscopy (AFM) measurement. The current–voltage (IV) measurements of the devices were performed in the probe station with Keithley 4200 Parameter analyser at room temperature under dark and in ambient conditions unless mentioned.

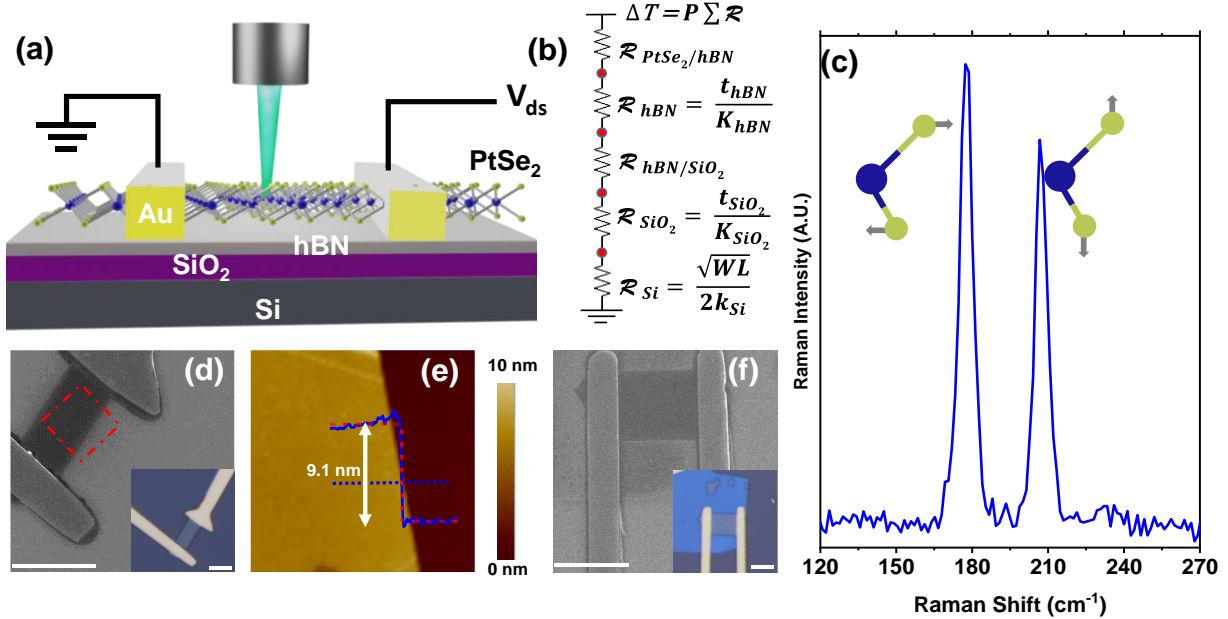

FIG.1. (a) The schematic diagram of PtSe$_2$ device substrate and the experimental I-V and Raman measurement setup. (b) The circuit diagram indicating the thermal resistance at different layers of the device. (c) The typical Raman spectrum of the PtSe$_2$ showing the characteristic E$_g$ and A$^1_g$ vibrational modes. (d & f) The SEM image and optical image (inset) of the PtSe$_2$ devices on SiO$_2$ and h-BN substrate respectively. The scale bar is 10 µm. (e) The AFM height image of the marked region of the PtSe$_2$ device in 1(d).

**Raman Thermometry**

To measure the rise in temperature of the device with the application of electric field, we have used the Raman spectrometer as a thermometer. The temperature-peak shift calibration factor of PtSe$_2$ is estimated from the temperature dependent Raman spectra. The Raman measurement were carried out in the Renishaw In via Raman spectrometer using 532 nm laser source under 50x lens and 30 µW laser power. The minimum possible laser power was used to avoid the laser induced heating of the sample. The heating of the substrate was achieved using the Linkum stage during the temperature dependent Raman spectra measurements. The bias voltage dependent Raman measurements were carried out in our home made probe setup under the Raman spectrometer and Keysight B2902A SMU. The voltage bias is carefully chosen in such a way that, we could able to record the Raman shift of the channel just before the breakdown voltage. The temperature-peak shift calibration factor is used to the estimate the device temperature at the corresponding voltage bias.

**Results and Discussions**

The schematic diagram of the PtSe$_2$ FET device on h-BN/SiO$_2$/Si substrate and experimental setup is shown in the FIG. 1a. The typical Raman spectra of PtSe$_2$ flake is shown in the FIG.1c, where the two most prominent peak E$_g$ and A$^1_g$ observed at ~178 cm$^{-1}$ and ~206 cm$^{-1}$ vibrated in-plane and out-of-plane direction respectively[18]. The Raman shift in both of the characteristic vibrational mode are tracked as a parameter to estimate the temperature in the Raman thermometry[19,20]. Before proceeding for the electro-opto-thermal measurements, we have performed the room temperature electrical characterization of our PtSe$_2$ FETs. As the thickness of the PtSe$_2$ (shown in FIG. 1e) considered in this experiment are in semi-metallic regime, there is no gate dependence observed in any of our devices[21]. Hence all the measurements are done at zero gate bias throughout the experiment.

The non-destructive Raman thermometry has been widely used to measure the local temperature of 2D materials[3,4,17,20,22]. To calibrate the Raman thermometer for measuring the local temperature of the $PtSe_2$, we have measured the temperature dependent Raman spectra of $PtSe_2$ from room temperature to 500 K. The lowest possible laser power is used to avoid the laser induced heating. The characteristic peak positions were obtained by the Lorentzian peak fitting to the spectra. The red shift of both the $E_g$ and $A^1_g$ peaks are observed as shown in the FIG. 2a and the first order temperature coefficient of Raman shift is found to be -0.017 ± 0.0007 $cm^{-1}.K^{-1}$ for $E_g$ and -0.010 ± 0.0005 $cm^{-1}.K^{-1}$ for $A^1_g$ as shown in the FIG. 2e.

Then, we have measured the bias dependent Raman spectra of the $PtSe_2$ on both $SiO_2$ and $h$-BN substrate as shown in the FIG. 2b-c. The applied bias in the devices is slowly increased so that we can able to measure the Raman spectrum just before the breakdown of the devices. We have waited sufficient time after applying the electric field and then recorded the Raman spectrum to ensure the proper heat spreading in the sample. At $V_D = 0$, we didn't see any change in the Raman spectrum of both $PtSe_2$ on $SiO_2$ and $h$-BN substrate. As the field increases gradually, a red shift in both the Raman modes is observed along with a slight increase in Raman intensity. The softening of the Raman mode is due to the current induced joule heating in the devices. From the amount of Raman shift, we have estimated the local temperature of the $PtSe_2$ layer by using the first order temperature coefficient of Raman shift obtained from the temperature dependent Raman measurement (shown in FIG. 3e). The temperature of $PtSe_2$ devices on both $SiO_2$ and $h$-BN substrate at each applied bias from 0 V up to breakdown voltage have been measured from the shift of both $E_g$ and $A^1_g$ peaks (shown in FIG. 3f-g). The temperature obtained from the Raman shift of both the peaks agree well with each other while $E_g$ peak is more sensitive to temperature than $A^1_g$ peak (showing a higher peak shift value) and hence the shift in the $E_g$ mode is considered for estimating the local temperature of the $PtSe_2$ layer throughout the experiment.

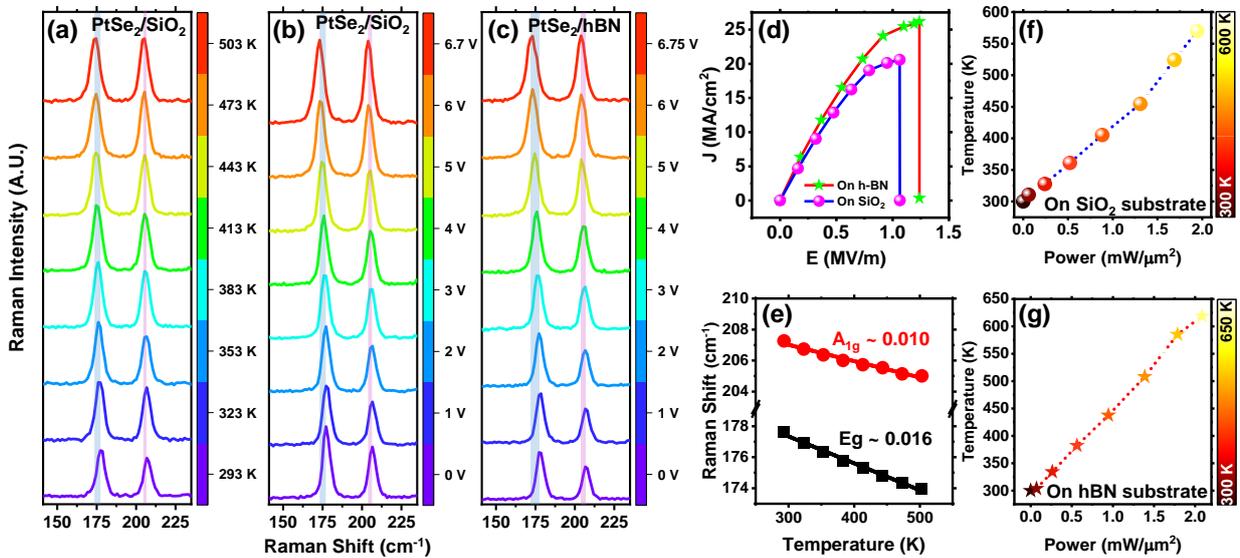

FIG.3. (a) Temperature dependent Raman spectra of $PtSe_2$. (b-c) The voltage bias dependent Raman spectra of $PtSe_2$ device on $SiO_2$ and h-BN substrate. (d) Current density verses applied electric field of both the devices. (e) Raman shift vs. temperature with linear fit to $E_g$ and $A^1_g$ peak. (f-g). Temperature extracted at each input electrical power from $E_g$ (solid line) and $A^1_g$ (dotted line) on both the substrates.

It has seen that, the $PtSe_2$ device on $SiO_2$ substrate is able to sustain up to 570 K before breakdown while the break down temperature of the device on $h$-BN substrate is 618 K. To understand the heat spreading mechanism in both the devices, we have estimated the ITC of $PtSe_2/SiO_2$ and $PtSe_2/h$-BN interface. The thermal healing length of $PtSe_2$ in the lateral direction is reported to be around 100 nm[17] but in our case the device channel length is more than 4 μm, so the heat dissipation into the contacts is ignored. Also the heat dissipation to the air is being ignored by considering air as an adiabatic medium with very low thermal conductivity[23]. The heat dissipation from channel into the substrate is the only pathway considered for the heat spreading. The FIG.1b shows the thermal resistance network for the heat dissipation into the substrate which define the total thermal resistance as the sum of the thermal resistance of each constituent layers and the interfaces as shown in equation.1 [17,20]. The out of plane thermal conductivity of $h$-BN ($K_{h-BN}$) as 5 W.m$^{-1}$.K$^{-1}$[24] and for $h$-BN/SiO$_2$ interface as 0.34×10$^{-7}$ K.m$^2$.W$^{-1}$ is being used in the calculation[25]. The thermal resistance at Si/SiO$_2$ has been neglected as it is extremely low at room temperature and the thermal properties of Si and SiO$_2$ is thoroughly studied[20,26]. The total thermal resistance is defined as change in the temperature divided by total power dissipated per unit channel area $R_{th} = \frac{\Delta T}{P_c}$. The electrical power dissipated per unit area in the channel after subtracting the contact resistance contribution is defined as $P_c = \frac{I^2R - 2I^2R_c}{A}$, where A, is the area of the channel, I is the current, R is the measured resistance and $R_c$ is the contact resistance. By plugging in each resistance contribution into the equation.1, the interfacial thermal resistance ($R_{int}$) of desired interface has been estimated. The reciprocal of $R_{int}$ gives the ITC value of the interface.

$$\mathcal{R} = \frac{\Delta T}{P_c} = \mathcal{R}_{PtSe_2/h-BN} + \mathcal{R}_{h-BN} + \mathcal{R}_{h-BN/SiO_2} + \mathcal{R}_{SiO_2} + \mathcal{R}_{Si/SiO_2} + \mathcal{R}_{Si} \ldots\ldots\ldots\ldots (1)$$

The ITC of PtSe$_2$/SiO$_2$ is estimated to be 14.2 MW.m$^{-2}$.K$^{-1}$ which agrees well as per reported by Liu et.al.[17], whereas the ITC of PtSe$_2$/$h$-BN interface is found to be 30.5 MW.m$^{-2}$.K$^{-1}$. The two fold increase in the ITC value of PtSe$_2$/$h$-BN interface is attributed to the ultra-smooth interface and high thermal conductivity of the substrate. Also, it is evident from the current density verses electric field plot that, the device on the h-BN substrate exhibits larger maximum power density and electric field persistence than the device on SiO$_2$ substrate at both room and low temperature (FIG. 3d, 4c). The break down current density of PtSe$_2$ on SiO$_2$ is 20.5 MA.cm$^{-2}$ while it increased to 26.1 MA.cm$^{-2}$ for PtSe$_2$ on h-BN substrate. The higher value of breakdown current density of the PtSe$_2$ on h-BN substrate is due to the efficient heat dissipation of PtSe$_2$ into the $h$-BN substrate owing to the following reasons. (i) The inertness and flatness of $h$-BN substrate as compared to the SiO$_2$ facilitating smooth heat conduction due to the reduction of interface acoustic phonon scatterings. (ii) The optical phonon energy of h-BN is two-fold higher than that of SiO$_2$ leading to better thermal coupling with PtSe$_2$.[] (iii) The thermal conductivity of $h$-BN is two order higher than that of SiO$_2$ causing better heat spreading in the $h$-BN substrate.[]

**Low Temperature transport in PtSe$_2$**

To understand the carrier transport mechanism and the breakdown in PtSe2 on both substrates, the IV characteristics of PtSe$_2$ up to breakdown have been measured at both room temperature and low temperature. The J-E plot of the PtSe2 at low temperature has been shown in the FIG.3 (a). There is a significant difference has been observed between the breakdown behavior of the devices at low temperature and room temperature. Typically the current-voltage at room temperature goes as linear and then saturate due to self-heating and then breakdown occurs. Meanwhile, at low temperature, in addition to the linear and saturated regime there is a super-linear region has been observed prior to breakdown as marked in light-pink color in the FIG.3 (a). Such super-linear behavior in IV at high applied electric field can be due to various reasons such as (i) the high contact resistance of the devices, (ii) the trap-assisted hopping conduction in the channel, and (iii) the charge carrier multiplication in the channel at high electric field.

When the contact resistance of a device is very high, the high applied electric field can lead to the local heating of the device. This thermal effect can improve the contact resistance of the device leading to a super-linear current at high bias. To understand the role of the contact resistance in out PtSe2 devices, we have fabricated the TLM-based PtSe$_2$ devices with different channel lengths (1 - 4 µm) as shown in the inset of FIG. 3b. The IV characteristic (at Room Temperature) of the devices with different channel length is shown in the FIG. 3b. In FIG. 3c, the width normalized total resistance vs. channel length (L) at different temperature are plotted and the linear fitting ($RW = R_s L + 2R_c$) is used to estimate the sheet resistance ($R_s$) and the contact resistance ($R_c$), where, R is the measured resistance, W and L are the channel width and length respectively. There is a linear temperature dependence behavior of $R_s$ and $R_c$ is observed. The contact resistance ($R_c$) is found to be in the range of 0.14-0.24 kΩ.µm and the $R_s$ is found to be in the range of 1.5-3.8 mΩ.cm for the probed devices in the temperature range of 150-460 K for the Ti 20 nm/Au 20 nm contact. As the contact resistance of our devices is in low regime, it is reasonable to say the device behaviour obtained in our measurements are not contact limited and the contact resistance does not contribute to the super-linear behavior in the IV.

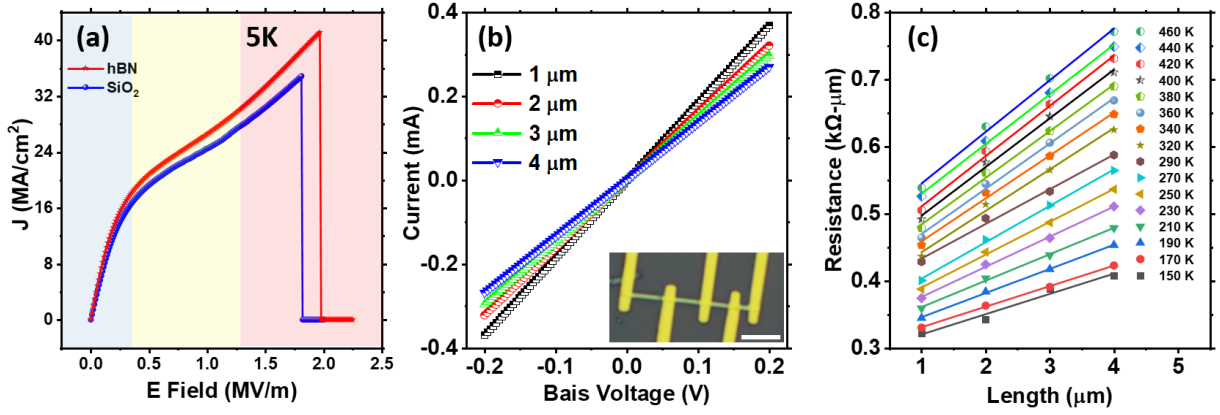

Fig.3- (b) IV characteristic of TLM at different channel length. The inset shows the OM image of the measured TLM devices with a scale bar of 5 µm (c) The total width normalized resistance vs. channel length at different temperature 150-460 K.

When there is trap states have the charge carrier in the channel material, the application of electric field can de-trap the carrier and contribute to the conduction by enhancing the conductivity. But such hopping transport only occurs at low applied electric field suggesting the observation of super-linear current at high applied field is not contributed by the trap-assisted hopping transport[27]. Furthermore, the hopping transport is dominant in the semiconductors, while the crystalline semimetallic PtSe$_2$ exhibits strongly interacting energy bands throughout the Brillouin zone and the presence of various electron and hole pockets around the Fermi level (As shown in SI FIG.xx) does not supports hopping transport as a dominant carrier transport mechanism[28].

To get the further insights on the carrier transport mechanisms in PtSe$_2$, we have fabricated the PtSe$_2$ Hall bar devices as shown in the FIG.4b. The two probe and four probe resistance vs temperature (RT) have been measured up to 2K. The increase in resistance with the increase in temperature suggest the semimetallic behavior of our few-layered PtSe$_2$ samples. The typical metallic/semimetallic RT behavior can be understood via the Bloch-Grüneisen (BG) equation as shown below.

$$R = R_0 + A_p \left\{\frac{T}{T_D}\right\}^p \int_0^{\frac{T_D}{T}} \frac{x^p}{(e^x-1)(1-e^{-x})} dx \ \ldots\ldots\ldots\ldots\ldots\ldots (1)$$

Here, $R_0$ is the residual resistance, $T_D$ is the Debye temperature, $A_p$ is the weight parameter and p is the power-law exponent, which is 2 for electron-electron (e-e) interactions, 3 for magnon interactions and 5 for electron-phonon (e-p) interactions.

The measured RT behavior of our few-layered PtSe$_2$ sample is shown in FIG. 4c. The positive residual resistivity ratio (RRR ~ $R_{300K}/R_{2K}$ ~ 3.1) and increase in resistance of the device with temperature (T > 10 K) signifies the semimetallic behavior of few layer PtSe$_2$ where the carrier transport occurs from band to band, showing no evidence of hopping or activation[28]. The FIG. 4c shows the measured RT curve fitted with the BG equation with R-square ~ 0.99 signifying a very good fitting of our data with the BG model. $R_0$, $A_p$, $T_D$ and p were considered as free fitting parameter while fitting the RT data. We have also measured various devices in both two-probe and four-probe configuration (Shown in SI FIG. 1) and the fitting parameters of the devices are tabulated in the Table. 1.

| Table.1 - The fitting parameters of R vs. T with BG equation. | | | | |
|---|---|---|---|---|
| Device name | Dimensions L μm × W μm × t nm | Contact configuration | $T_D$ | p |
| D1 | 6.2 × 3 × 21 | 2 probe | 482 ± 3.6 | 1.97 ± 0.01 |
| D2 | 7.3 × 1.3 × 10 | 2 probe | 363 ± 3.6 | 2.18 ± 0.02 |
| D3 | 1.8 × 0.6 × 5 | 2 probe | 557 ± 8.6 | 1.99 ± 0.03 |
| D4 | 9.3 × 6.3 × 10 | 4 probe | 510 ± 4.2 | 2.17 ± 0.02 |
| D5 | 13 × 6.5 × 30 | 4 probe | 439 ± 5.1 | 2.27 ± 0.03 |

We can see from the Table.1 that, the best fit of our R vs. T data is occurring with the power exponent value as ~ 2. This signifies the dominant carrier interaction mechanism to be the e-e interaction in $PtSe_2$ throughout the thickness range probed here. The similar e-e interaction dominant behavior has been reported in other topological semimetal like $WTe_2$ and $PtTe_2$ etc[28,29]. Furthermore, the Debye temperature for $PtSe_2$ is found to be in the range of 360-560 K in our devices.

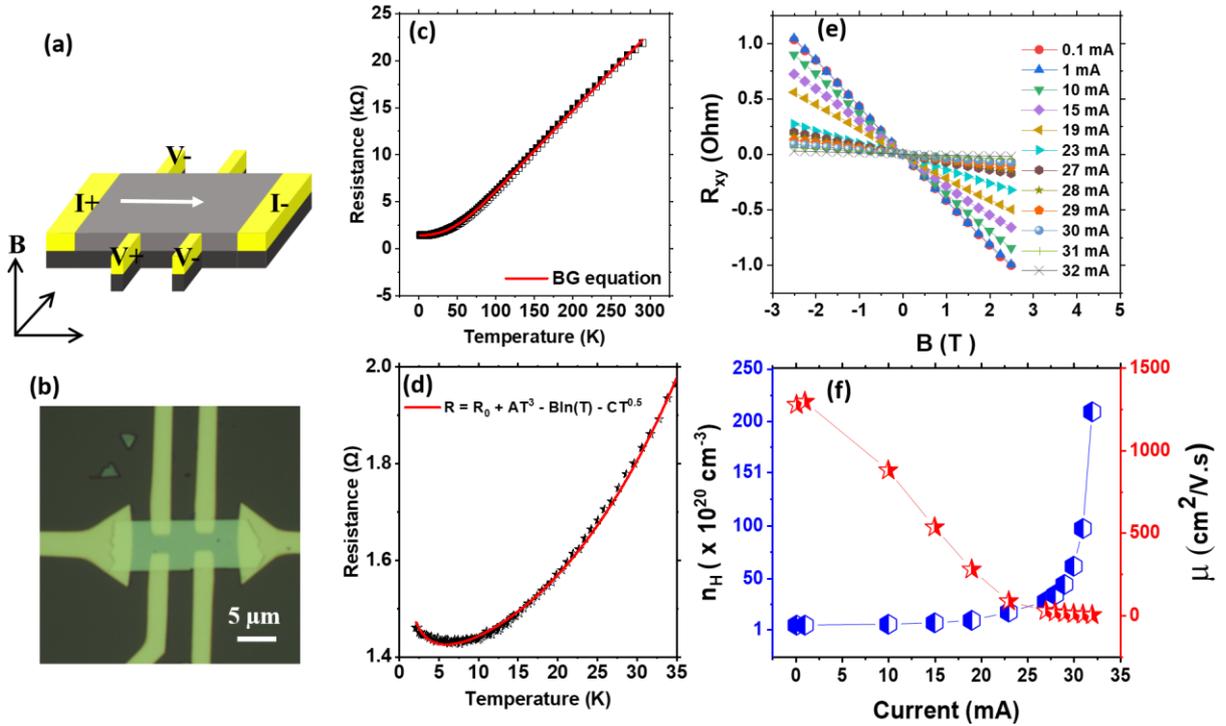

Fig.4- (a) The schematic representation of the devices with electrical connections to the electrodes. (d) The OM image of the Hall-bar device measured. (b) The R-T curve fitted with the BG equation. (e) The low temperature resistance up-turn fitted with the model. (c) The transverse resistance vs magnetic field measured at different applied current. (f) The Hall carrier concentration and the mobility extracted from the Hall resistance measurement.

The low temperature resistance up-turn behavior has been observed in out RT data below 10 K temperature as shown in FIG. 4d suggesting the quantum correction to the resistance as origin. Such quantum corrections at low temperature can be due to (i) Weak Localization (WL), (ii) Kondo effect, and (iii) e-e interaction. The WL correction to resistance arises due to the constructive interference of electron wave function (coherent back scattering) resulting the rise in resistance at low temperature. By the application of external magnetic field the coherence can be broken and the resistance up-turn vanishes. We rule out the WL as the cause of the resistance up-turn as it is still there even in applied magnetic field as shown in SI FIG. XX. The Kondo effect arises due to the scattering of conduction electron with the local magnetic moment of magnetic impurities. The possibilities of Kondo effect as the correction to the resistance at low temperature has been eliminated due to the absence of magnetic impurities in our PtSe$_2$ devices. Hence the e-e interaction is expected to be the cause of the resistance up-turn in our devices, which need to be examined. The e-e interaction to the resistance in metal has been proposed to be ΔR α –ln(T) in 2D material and ΔR α –T$^{0.5}$ for 3D metals[28]. So we have fitted the resistance with following equation,

$$R = R_0 + AT^3 - Bln(T) - CT^{\frac{1}{2}} \ldots \ldots \ldots \ldots \ldots \ldots (2)$$

where, the parameter A is contributed from e-ph scattering, B and C gives the contribution of 2D and 3D e-e scattering respectively. The up-turn resistance fitted with the equation-2 is shown in FIG.4d. It is noticed that the up-turn resistance has been fitted well with the dominant value of the coefficient of –ln(T) confirming the e-e interactions to be governing carrier transport mechanism in PtSe$_2$.

To understand, the roles of the carrier multiplication in PtSe$_2$ channel, the transverse Hall resistance with magnetic field sweep have been measured at each applied electric field in all the linear, saturated and super-linear regime of current at 5K temperature (as shown in FIG.4e). The linear behavior of transverse resistance with applied electric field signifies the role of single type of carrier dominating the transport in low temperature in PtSe$_2$, which has been previously observed by Bonell et.al[30]. Furthermore, the negative Hall resistance confirms the electrons as dominating carrier in PtSe$_2$ at the low temperature.

The carrier concentration ($n_H$) and mobility ($\mu_H$) have been extracted from the Hall resistance data using $n_H = \frac{1}{e} \frac{dB}{dR_{xy}}$, and $\mu_H = \frac{1}{R_{xx} \times n_H \times e}$ respectively (as shown in FIG.4f). Here, we can clearly see the Hall carrier concentration is almost constant in the linear and saturated region while, it suddenly rises as we further increase the current. While the Hall carrier concentration increases, there as a gradual fall in the Hall mobility of PtSe$_2$ with the applied current. Such fall in the mobility with applied current is mainly due to the rise in the local temperature of the PtSe$_2$ channel due to the Joule heating of the devices. The multi-fold enhancement of the Hall carrier concentration confirms the hot carrier multiplication causing the super-linear current in PtSe$_2$ before the breakdown at low temperature. Hence the breakdown of PtSe$_2$ at low temperature is mainly due to the hot carrier multiplication.

**Conclusions**

In conclusions, we have explored the major difference between the electrical transports in PtSe$_2$ at low/ high temperature/electric field up to breakdown. The room temperature breakdown of the device is mainly due to the self-heating of the device at high applied current. The heat dissipation of the device has been understood with the help of Raman thermometry. The ITC of PtSe$_2$/SiO$_2$ interface is found to be ~ 14.2 MW.m$^{-2}$.K$^{-1}$ while there is a two-fold enhancement in the ITC value for the PtSe$_2$/*h*-BN interface confirming the impact of dielectric environment in the performance of the device. The better heat dissipation at PtSe$_2$/*h*-BN interface is mainly due to the ultra-clean 2D-2D interface and the high thermal conductivity of the *h*-BN in comparison to SiO$_2$, spreading the heat quickly from the active region of the device into the substrate. The super-linear behavior of the current prior to the breakdown of PtSe$_2$ at low temperature is mainly due to the hot carrier multiplication as confirmed via Hall measurements. The understanding of breakdown mechanism at both low/high temperature along with the thermal management in PtSe$_2$ device pave the way for designing high performance, energy efficient electronic devices, operating at extreme conditions.

See the supplementary material for the OM, AFM and TEM images, I-V analysis plots and the details of computational analysis.

We acknowledge the financial support from the Ministry of Human Resource Development (MHRD), the Government of India (GOI) via STARS grant [STARS/APR2019/NS/631/FS] and IIT Madras for setting up research centre "Centre for 2D Materials Research and Innovations" through Institute of Eminence scheme. B.B acknowledge Mr. Samim Hossain for the help in collecting the AFM images. We also acknowledge the Material Science Research Centre (MSRC) IIT Madras for providing e-beam deposition facility.**Conflict of Interest**

The authors have no conflicts to disclose.

**DATA AVAILABILITY**

The data that support the findings of this study are available within the article and its supplementary material.